\documentclass[submission,copyright,creativecommons]{eptcs}
\providecommand{\event}{HCVS 2018}

\usepackage{comment}
\usepackage{amssymb}
\usepackage{amstext}
\usepackage{amsmath}
\usepackage{latexsym}
\usepackage{qtree}
\usepackage{ifsym}
\usepackage{verbatim}
\usepackage{url}
\usepackage{proof}
\usepackage{listings}
\usepackage{color}
\usepackage{fancyvrb}
\usepackage{paralist}
\usepackage{todonotes}

\usepackage{relsize}
\usepackage{float}
\floatstyle{boxed}
\newfloat{boxedfg}{thp}{lop}[figure]
\floatname{boxedfg}{Figure}

\usepackage{enumitem}

\usepackage{mathptmx}
\usepackage[all]{xy}
\usepackage{comment}
\usepackage{multirow}

\newtheorem{example}{Example}

\title{Towards Coinductive Theory Exploration in Horn Clause Logic: Position Paper \vspace*{-0.2in}}
\author{Ekaterina Komendantskaya
\institute{School of Mathematical and Computer Sciences\\ Heriot-Watt University,  UK}
\email{ek19@hw.ac.uk}
\and
Yue Li
\institute{School of Mathematical and Computer Sciences\\ Heriot-Watt University,  UK}
\email{yl55@hw.ac.uk}
}

\def\titlerunning{Coinductive Theory Exploration in Horn Clause Logic}
\def\authorrunning{Komendantskaya \& Li}

\begin{document}
\maketitle
\begin{abstract}
  Coinduction occurs in two guises in Horn clause logic: 
	in proofs of self-referencing properties and relations, and in proofs involving construction of (possibly irregular) infinite data.
	Both instances of coinductive reasoning appeared in the literature before, but a systematic analysis of 
	these two kinds of proofs and of their relation was lacking. 
  We propose  a  general proof-theoretic framework for  
	handling both kinds of coinduction arising in Horn clause logic.
	To this aim, we propose a coinductive extension of Miller et al's framework of \emph{uniform proofs}  and prove its soundness
	relative to coinductive models of Horn clause logic. % of the resulting calculi.
\end{abstract}

%\begin{keyword}
% Horn Clauses, Coinduction, Corecursion, Coinductive Invariants
%\end{keyword}

%\end{frontmatter}

\section{Problem Statement}\label{sec:intro}

%\subsection{Motivation}

Coinductive proof methods have seen major developments in the last decade, and are reaching the point of maturity
when coinductive proofs are used and implemented on par with inductive proofs. 
This step-change is facilitated by results from several research areas: coalgebra, 
fixed point theory, type theory, proof theory, automated deduction.
In this abstract, we discuss a new coinductive approach to Horn clause logic.
%Of course results in these areas have subtle connections, that are also being actively explored~\cite{B18,BasoldG16,KomendantskayaP16}.

%For simplicity, we will assume that $\Gamma$ is moreover 
A Horn clause fragment of FOL, named \emph{fohc}, is given by the following syntax:
\begin{gather*}
D \ ::= A \mid G \supset D \mid  D\land D  \ \mid \forall Var \ D\\
G \ ::= \top   \mid G \land G \mid  G\lor G  \mid \exists Var \ G
\end{gather*}
where $A$ stands for the set of atomic first-order formulae of a given signature, and $D$ and $G$ -- for sets of definite Horn clauses and definite Horn goals, respectively. A  \emph{theory} $\Gamma$ is a set of $D$-formulae.

% Restricting our attention to theories expressed in \emph{fohc} will allow us to rely on well-studied operational semantics (resolution) and fixed point (Knaster-Tarski) semantics of Horn clause logic, while still working with sufficiently expressive (in fact, Turing complete) language.

%Many coinductive calculi, and many algorithms of automated inference in first-order logic rely on a variant of the L\"{o}b rule~\cite{Sol76}:
%\[\infer[\textit{L\"{o}b}]{\Sigma;\Gamma \longrightarrow F}{\Sigma;\Gamma, \lhd F \longrightarrow  F}\]
%where $\Gamma$ is a  (first-order) theory and $F$ -- a (first-order) formula. Intuitively, $\lhd F $ 
%is taken as a coinductive hypothesis, that can be applied later in the proof. 
%Soundness of cycle-detection algorithms like  CoLP~\cite{GuptaBMSM07} usually relies on soundness of the L\"{o}b rule. 

First coinductive interpretation to Horn clause logic was given by Apt and van Emden in the 80s: 
The \emph{greatest complete Herbrand model} for a
theory $\Gamma$ is the largest set of finite and infinite
ground terms \emph{coinductively entailed} by $\Gamma$'s clauses.
%
%\knote{give the DF}

\begin{example}
  Consider the three Horn clause theories $\Gamma_1$, $\Gamma_2$ and $\Gamma_3$ in Table~\ref{tab:3}.
%  $\Gamma_1 = \{ 1.\ \forall x, p(x) \supset p(x) \}$, \\ $\Gamma_2 = \{ 2.\ \forall x, p(f\ x) \supset p(x)  \}$,\\
%  $\Gamma_3 = \{ 3.\ \forall x, p( x) \supset p(f\ x)\}$. 
		None of them has a meaningful inductive interpretation. However,
   they all have  greatest (complete) Herbrand models, as Table~\ref{tab:3} shows. 
	These models define their coinductive interpretation. Notice how, depending on the clause structure, the models will differ:
	they may be given by finite sets of finite atomic formulae (for $\Gamma_1$),  or infinite sets of finite and infinite formulae ($\Gamma_2$), or finite sets of infinite formulae ($\Gamma_3$).
	Note that $\Gamma_3$ is a prototypical example of a productive stream definition~\cite{KL17}: just substitute $f$ by a stream constructor $cons(a,\_)$ to obatin a definition of the infinite stream of $a$'s. Only one infinite term
	satisfies $\Gamma_3$.
	   \end{example}

\begin{table}
{\small{
	%\centering
\begin{tabular}{|p{2.7cm} || p{2.2cm} | p{6cm} | p{2.7cm}|}
\hline
hohc theory: &  $\Gamma_1:$ & $\Gamma_2:$ & $\Gamma_3:$   \\
 & $1.\ \forall x\ p(x) \supset p(x)$ & $2.\ \forall x\ p(f\ x) \supset p(x)  $ &
 $3.\ \forall x\ p( x) \supset p(f\ x)$\\ \hline  
greatest complete  Herbrand model: &   $ \{\mathtt{p(a)}\}$ & $ \{\mathtt{p(a),}$ $\mathtt{ p(f(a)), p(f(f(a))},$ $\ldots , \mathtt{p(f(f\ldots)}\}$  & $\{\mathtt{p(f(f\ldots)}\}$  \\    
%For query $ \mathtt{p(X)}$, computes the answer: & ~~$\{\mathtt{X} \mapsto \mathtt{f (f ...)}\}$ & ~~$\textit{id}$ & ~~$\textit{id}$ \\ 
%For query $ \mathtt{p(a)}$: &  fails & computes  ~~$\textit{id}$  & computes ~~$\textit{id}$   \\
\hline
\end{tabular}\caption{\textbf{Examples of greatest (complete) Herbrand models for fohc theories $\Gamma_1$, $\Gamma_2$, $\Gamma_3$.} 	We add an arbitrary constant symbol $a$ to the signature, in order to have ground instances of formulae in the models.}\label{tab:3}}}
\end{table}

It has always been problematic to match the greatest complete Herbrand models with equally rich operational semantics. 
It is long known that infinite (SLD)-resolution derivations correspond to coinductive models~\cite{Llo88}.
Some infinite derivations may be terminated if a loop invariant (also known as \emph{coinductive invariant}) is found.
%Consider the following examples.
%For example, one practical problem arises in implementation of coinductive proofs:
%that arises in proof theory and automated deduction of coinductive properties:
The problem is then to automate the discovery of coinductive invariants. 
%To illustrate the problem, consider the following simple example.
To illustrate how difficult this may prove to be, consider the following example.
Given our three theories $\Gamma_1$, $\Gamma_2$ and $\Gamma_3$, suppose we want to prove a property $p(a)$ by coinduction. % (here, $p$ is an  arbitrary predicate, and $a$ -- an arbitrary constant).
%Consider the following example.

%Our key observation is that shapes of coinductive invariants required to coinductively prove atomic formulae in \emph{fohc} can vary so significantly that they may not be even expressible in the syntax of \emph{fohc}.

%\ynote{edited this sentence. The old sentence has ``on the one hand... secondly...'' structure. Now we only have one aspect to say, i.e. inadequacy of fohc, and we now do not mention hohh, so there is no second aspect.}

%Our key observation is that, on the one hand, shapes of coinductive invariants required to coinductively prove atomic formulae in \emph{fohc} can vary significantly and  secondly that they may not be even expressible in the syntax of \emph{fohc}.

%, but on the other hand, such diversity can be captured using the \emph{higher-order hereditary Harrop logic} (\emph{hohh}). \emph{Hohh} turns out to be a versatile\footnote{For traditional applications of \emph{hohh}, see \cite{MN12}} logic suitable as a meta-language for coinductive theory exploration in Horn clause logic.  We expose these points in the following examples.

%\ynote{Added the sentence above to highlight the main discovery of the paper.}

%\noindent To prove $\Gamma \vdash p(a)$, we may try to apply the L\"{o}b rule, and take $p(a)$ as a coinductive hypothesis. 
%However, in a goal-directed proof construction, it is actually a matter of luck whether such a strategy will work. 

\begin{example}\label{ex:1.1}
%If we are lucky, the context may be given by:
%Suppose $\Gamma_1 = \{ 1.\ \forall x, p(x) \supset p(x)$ \}, and we wish to prove $p(a)$.
%and  the  coinductive proof for $p(a)$ will go through by applying the L\"{o}b rule directly. Here is its schematic representation (we assume some standard basic rules that work with quantifiers and implications, they will be made precise in the technical sections). 
For $\Gamma_1$, we will observe the following resolution steps:
%Underlined is the step where the coinductive assumption $CH_1 = p(a)$ is added to the context $\Gamma_1$:
\[p(a) \stackrel{apply\ 1}{\longrightarrow} p(a) \stackrel{apply\ CI_1}{\longrightarrow} \checkmark\]
Clearly, $p(a)$ is the coinductive invariant (denoted as $CI_1 = p(a)$), the derivation is cyclic, and we can terminate soundly by noting this fact. Note how $\Gamma_1$'s model  in Table~\ref{tab:3} agrees with this conclusion.
Coinductive logic programming (CoLP)~\cite{GuptaBMSM07} handles such cases well: its method of loop detection is able to
find that $p(a)$ is looping and thus find the correct coinductive hypothesis. 
\end{example}

However, it is entirely possible that an environment $\Gamma$ entails $p(a)$, 
%yet the coinductive invariant in the $p(a)$'s derivation is different from $p(a)$. In this case, taking $p(a)$ as a coinductive hypothesis will not result in a successful proof.
yet $p(a)$ does not occur as an invariant in its infinite derivation.

\begin{example}\label{ex:2 new}
Consider $\Gamma_2$.
%is exactly such a case. 
Trying to replicate the coinductive proof of Example~\ref{ex:1.1} with coinductive invariant $p(a)$  would not work, as the coinductive invariant will not apply at any stage (the derivation does not have cycles):
\[p(a) \stackrel{apply\ 2}{\longrightarrow} p(f \ a) 	\stackrel{apply\ 2}{\longrightarrow} p(f \ (f\ a))  \longrightarrow \ldots \]
%Underlined is the step where the coinductive assumption $CH_1 = p(a)$ is added to the context $\Gamma_2$, but it will not be useful in this derivation. 

A valid (as well as useful) coinductive invariant in this proof is $CI_2 = \forall x\ p(x)$. So, \emph{given a suitable calculus}, we can first coinductively prove $\Gamma_2 \vdash \forall x\ p(x)$, and then obtain $\Gamma_2 \vdash p(a)$ as a corollary. Note, however, that the formula $\forall x\ p(x)$ does not satisfy the syntax of a goal formula in \emph{fohc}. And note also that loop-detection methods like  CoLP~\cite{GuptaBMSM07} cannot handle such cases:  no loop (i.e. no unifying subgoals) can be found in this derivation.

Generally, discovering a suitable coinductive invariant may be a difficult task. Consider the following example, inspired by a similar example in~\cite{FKS15}.
%So, the suitable calculus  for suCI a proof is not fohc. 

%$\underline{\forall x, p(x)} \stackrel{\forall-I }{\longrightarrow} p(a) \stackrel{apply\ 2}{\longrightarrow} p(f\ a) \stackrel{apply\ CH_1}{\longrightarrow} \checkmark$.
\end{example}

\begin{example}\label{ex:3 new}
	Suppose we want to prove $p(a)$ given the theory $\Gamma_4:$\\
	$4.1.\ \forall x\ p(f\ x) \land q(x) \supset p(x)$\\
	$4.2.\ q(a)$\\
	$4.3.\ \forall x\ q(x)\supset q(f\ x)$

        Its greatest complete Herbrand model is given by:

        $$ \mathtt{ \{p(a), p \ (f\ a), p (f\ (f\ a)), \ldots  , p(f(f(\ldots)))}$$ $$\mathtt{ q(a), q (f\ a), q (f\ (f\ a)), \ldots ,q\ (f(f(\ldots))) \ \}}$$
        Thus, $p(a)$ we seek to prove is coinductively valid.
        
	It will give the following resolution trace:
	\[p(a) \stackrel{apply\ 4.1}{\longrightarrow} p(f \ a) \land q(a) 	\stackrel{apply\ 4.2}{\longrightarrow} p(f \ a)  \stackrel{apply\ 4.1}{\longrightarrow} p(f \ f \ a) \land q(f\ a) 	\stackrel{apply\ 4.3}{\longrightarrow} \ldots  \]
The coinductive invariant $CI_1 = p (a) $ will not apply here, despite $p(a)$ being in the model of $\Gamma_4$.	Actually, neither  $CI_1 = p(a)$ nor $CI_2 = \forall x\ p(x)$ would work as a suitable coinductive invariant.  However, given a suitable calculus, we would be able to coinductively prove $ \Gamma_4 \vdash \forall x\ (q(x) \supset p(x))$, 
	from which $ \Gamma_4 \vdash p(a)$ can be proven as a corollary.
	Again, note that $CI_3 = \forall x\ (q(x) \supset p(x))$ cannot be a goal formula in \emph{fohc}, so we will need a different language for reasoning about coinductive invariant of the proof of $ \Gamma_4 \vdash p(a)$.
\end{example}

Finding a suitable coinductive invariant in a goal-directed proof search  may require coming up with recursive terms on top of finding a suitable shape for the coinductive invariant, as the next example shows:  

\begin{example}\label{ex:4 new}
Given a theory
$\Gamma_3$ from Table~\ref{tab:3}, % = \{ 4.\ \forall x, p( x) \supset p(f\ x)\}$ and a task to prove $\Gamma_4 \vdash \exists x, p(x)$,
the goal-directed search by resolution will result in a derivation:
\[\underline{p(x)} \stackrel{apply\ 3, [x \mapsto f(x_1)]}{\longrightarrow} p(x_1) 	\stackrel{apply\ 3, [x_1 \mapsto f(x_2)]}{\longrightarrow} p(x_2)  \longrightarrow \ldots \]
 None of the sub-goals can serve as a suitable coinductive invariant. The correct coinductive invariant in this derivation is
$ p (\textit{fix}\ \lambda x. f\ x)$, where the fixpoint term $\textit{fix}\ \lambda x. f\ x$ should be intuitively understood as a recursive definition for an infinite term $(f(f\ldots))$. 
Compare also with $\Gamma_3$'s model in Table~\ref{tab:3}, and its only inhabitant $\mathtt{p(f(f\ldots)}$.

Thus, we would like to  coinductively prove  $\Gamma_3 \vdash  p (\textit{fix}\ \lambda x. f\ x)$ in a suitable logic, and then get $\Gamma_3 \vdash \exists x, p(x)$ as a corollary.
Yet again,  $p (\textit{fix}\ \lambda x. f\ x)$ is not a formula of \emph{fohc}, because of the 
%: the reason this time is not its formula-level shape, but the higher-order 
syntax of $\textit{fix}\ \lambda x. f\ x$ is not in FOL. 
%\ynote{Minor rephrasing, for change see source code, the original words are commented out to avoid referring to the order problem at this early stage. }
\end{example}

\begin{figure}[t]
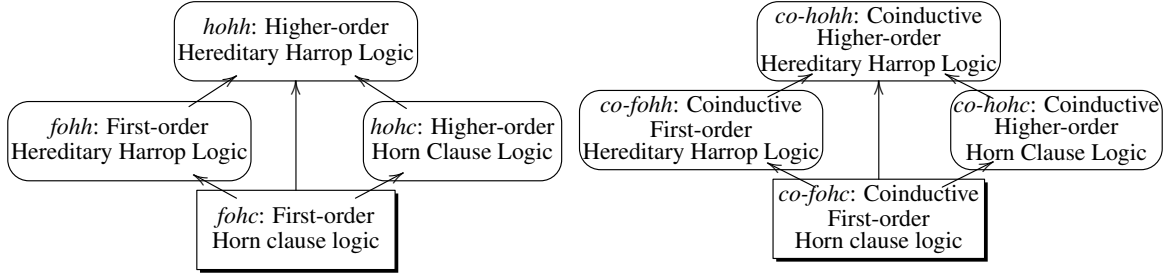

\footnotesize{
$$
\xy0;/r.035pc/: 
(0,0)*[o]=<92pt,30pt>\hbox{\txt{\emph{hohh}: Higher-order\\ Hereditary Harrop  Logic}}="d"*\frm<8pt>{-},
(-150,-90)*[o]=<92pt,30pt>\hbox{\txt{\emph{fohh}: First-order\\ Hereditary Harrop  Logic}}="c"*\frm<8pt>{-},
(150,-90)*[o]=<75pt,30pt>\hbox{\txt{\emph{hohc}: Higher-order\\ Horn Clause Logic }}="b"*\frm<8pt>{-},
(0,-170)*[o]=<75pt,30pt>\hbox{\txt{\emph{fohc}: First-order\\ Horn clause logic }}="a"*\frm<8pt>{-,},
"c";"d" **\dir{-} ?>*\dir{>},
"b";"d" **\dir{-} ?>*\dir{>},
"a";"d" **\dir{-} ?>*\dir{>},
"a";"c" **\dir{-} ?>*\dir{>},
"a";"b" **\dir{-} ?>*\dir{>}
\endxy
\ \ \ 
%$$
%{\footnotesize{
%$$
\xy0;/r.035pc/: 
(0,0)*[o]=<92pt,30pt>\hbox{\txt{\emph{co-hohh}: Coinductive \\ Higher-order\\ Hereditary Harrop  Logic}}="d"*\frm<8pt>{-},
(-160,-80)*[o]=<92pt,30pt>\hbox{\txt{\emph{co-fohh}: Coinductive \\ First-order\\ Hereditary Harrop  Logic}}="c"*\frm<8pt>{-},
(160,-80)*[o]=<80pt,30pt>\hbox{\txt{\emph{co-hohc}: Coinductive \\ Higher-order\\ Horn Clause Logic }}="b"*\frm<8pt>{-},
(0,-160)*[o]=<80pt,30pt>\hbox{\txt{\emph{co-fohc}: Coinductive \\ First-order\\ Horn clause logic }}="a"*\frm<8pt>{-,},
"c";"d" **\dir{-} ?>*\dir{>},
"b";"d" **\dir{-} ?>*\dir{>},
"a";"d" **\dir{-} ?>*\dir{>},
"a";"c" **\dir{-} ?>*\dir{>},
"a";"b" **\dir{-} ?>*\dir{>}
\endxy
$$}
%\vspace*{-0.1in}
\caption{\textbf{Left: uniform proof diamond by Miller et al~\cite{MN12}.} 
\textbf{Right: coinductive uniform proof diamond proposed in this paper.} The arrows show syntactic extensions from first-order to higher-order, from Horn to hereditary Harrop clauses.}\label{fig:diamonds}
	%\end{itemize}%}
\end{figure}

%In this paper, we take a first step towards a systematic methodology for discovery of suitable coinductive invariants in a goal-directed
%proof search in first-order logic.
Taking the assumption that a theory $\Gamma$ and a formula $F$ are expressed in \emph{fohc},
we can show that there are four different classes of coinductive proofs for $\Gamma \vdash F$, and they are all characterised by the
logic in which the coinductive invariant of the goal-directed derivation of $F$ can be expressed and proven.
%To obtain this main result, we make several technical contributions:  
%
%\begin{itemize}
%	\item 
	We take the uniform proofs of Miller, Nadathur et. al~\cite{MN12}, and in particular the four  uniform proof logics \emph{fohc}, \emph{fohh}, \emph{hohc}, \emph{hohh} (see Figure~\ref{fig:diamonds}), as a basis for our classification of the expressivity of the coinductive invariants. For example, coinductive invariant of Example~\ref{ex:1.1} belongs to \emph{fohc},  coinductive invariants of Examples~\ref{ex:2 new} and \ref{ex:3 new} -- to \emph{fohh}, and the coinductive invariant of Example~\ref{ex:4 new} -- to \emph{fohc} enriched with fixpoint terms.
	Horn clauses defining irregular streams will require the syntax of \emph{hohh} with fixpoint terms.
        \begin{example}\label{ex:from}
          Theory $\Gamma_5$ defines an infinite irregular stream
          $[0, (s\ 0), (s \ (s\ 0)), \ldots]$:

          $$5.1. \ \forall xy\quad \textit{from}\ (s\ x)\ y \supset  \textit{from}\ x\ (\textit{scons}( x , y))$$

          \noindent The infinite derivation for the above stream is  given by $$\textit{from}\ 0\ y \stackrel{apply 5, [y \mapsto \textit{scons}( 0 , y')]}\to \textit{from}\ (s\ 0)\ y' \to \cdots$$

          It cannot be handled by state-of-the-art coinductive theorem provers such as CoLP~\cite{SimonEtAl06,KL17}, as the method of loop detection fails for this example (the subgoals do not unify).
          
          \end{example}
        
          % \ynote{changed from co-hohh to hohh.}
         In the next section,  we will show that this example, too, can be handled by \emph{coinductive uniform proofs} and falls under the classification of Figure~\ref{fig:diamonds}.
This classification thus provides foundations for automated exploration of coinductive invariants for proofs with coinductive theories expressed in Horn clause logic.

\section{Meta-theory: Coinductive Uniform Proofs}
\label{sec: languages and rules}
Our term system extends simply typed lambda terms (typically $M,N$) by allowing constructs of the form $\textit{fix}\ \lambda x\,.\ M$, which shall satisfy standard guarding conditions %(the details of which  are not discussed in this abstract),
to denote infinite objects. We use $=_{\textit{fix}\beta}$ for equivalence of two infinite objects (formal details omitted). For example, for a regular fixed point term of Example~\ref{ex:4 new}:
\[\textit{fix}\ \lambda x\,.\, f\ x =_{\textit{fix}\beta} f(\textit{fix}\ \lambda x\,.\, f\ x) =_{\textit{fix}\beta} f(f (\textit{fix}\ \lambda x\,.\, f\ x)) =_{\textit{fix}\beta}\cdots \]

%One more example: concerning 
The infinite stream $[0,\, (s\ 0),\, s(s\ 0),\, s(s(s\ 0)), \ldots]$ %, the computed by goal $(\textit{from}\ 0\ y)$ w.r.t Horn clause \ref{prog: from}, we denote it 
is defined by the  higher-order term 
$$ (\left( \textit{fix}\ \lambda f n\ .\ \textit{scons}\ ( n,  \left(f\  \left(s\ n\right)\right)\right)) \ 0)$$ for which we write $\textit{fr\_str}\ 0$ as a short hand, and which satisfies the following relations\[\textit{fr\_str}\ 0 =_{\textit{fix}\beta} \textit{scons}( 0,  \left(\textit{fr\_str} \left(s\ 0\right) \right)) =_{\textit{fix}\beta} \    \textit{scons} \left( 0,  \left(\textit{scons}\left( \left(s\ 0\right),  \left(\textit{fr\_str} \left(s^2\ 0\right) \right)\right)\right)\right) =_{\textit{fix}\beta}\cdots \] 
The rest of syntax specifications follow the uniform proof theory. We use simple types involving the formula type $o$, and terms are built using constants from a signature $\Sigma$ and variables from the countably infinite set \textit{Var}. An atomic formula $B:o$ has the form $(h\ N_1\ \ldots \ N_n)$ where $h$ is either a constant different from $\land, \lor, \forall\!_\tau,\exists_\tau$ and  $\supset$,  or a variable; $B$ is \emph{rigid} (respectively,  \emph{flexible}) if $h$ is a constant (respectively, variable). A term is \emph{closed} if it does not have free variables. We use $\equiv$ for syntactical identity modulo $\alpha-$equivalence, $=_\beta$ for $\beta-$equivalence. We define $\mathcal{U}^\Sigma_1$ as the set of all terms over $\Sigma$ that do \emph{not} contain $\forall\!_\tau$ and $\supset$, and  $\mathcal{U}^\Sigma_2$ as the set of all terms over $\Sigma$ that do \emph{not} contain $\supset$. Table~\ref{tab:up} defines, for each of the four languages, the set $D$ of \emph{program clauses}  and the set $G$ of \emph{goals}. Given a signature $\Sigma$, a \emph{program} $P$ is a finite set of \emph{closed} $D$-formulae over $\Sigma$.

 We have two kinds of sequents. One kind of sequents are in the form $\Sigma;P\longrightarrow G$, encoding the proposition that the closed goal formula $G$ is 
provable in intuitionistic logic from the program $P$ on $\Sigma$. We use Miller et al's uniform proof rules (with slight extension to support the $=_{\textit{fix}\beta}$ relation, see Figure~\ref{fig: inductive uniform proof}) to prove sequents of this kind. We are interested in proving the other kind of sequents, which are in the form  $\Sigma;P\looparrowright  G$,  encoding that the closed goal formula $G$ is \emph{coinductively} provable from the  program $P$ on $\Sigma$. 

 Proving sequents on $\looparrowright$ is closely related to proving sequents on $\longrightarrow$, and for this point we give both formal and informal explanations.  Informally, consider the scenario where we begin with proving  $\Sigma;P\looparrowright  G$, which amounts to prove $\Sigma;P,G\longrightarrow  G$ next, but the way we can apply inference rules to prove $\Sigma;P,G\longrightarrow  G$  is more \emph{restricted}, compared to a related but different scenario in which we begin with proving $\Sigma;P,G\longrightarrow  G$. The  motivation for such restriction is to ensure consistency, i.e. to avoid erroneously making arbitrary formulae coinductively provable.  Formally, we use the \textsc{co-fix} rule (Figure~\ref{fig: general co-fix}) for sequents on $\looparrowright$,  and we introduce the notation $\langle \rangle$ in the \textsc{co-fix} rule, so that a formula marked with $\langle \rangle$ is \emph{guarded}\footnote{There are two distinct notions of \emph{guard} in coinductive uniform proof: one is for the syntax of fixed-point terms, to ensure that they model infinite objects; the other is for formulae in certain sequents, to ensure consistency.} and a sequent with guarded formulae shall be reduced using rules in Figure~\ref{fig: rules with angles}, which encodes the restriction we mentioned in the earlier informal account. 

A  (coinductive uniform) \emph{proof}  
is a finite tree such that the root is labeled with $\Sigma;P\looparrowright M$, and leaves are labeled with \emph{initial sequents} which are sequents that can occur as a lower sequent in the rules \textsc{initial} or $\textsc{initial}\langle\rangle$. A proof is constructed in \textit{co-fohc} if all formulae in the proof  satisfy the language syntax of \textit{co-fohc}.  Proofs constructed in \textit{co-fohh}, \textit{co-hohc}, or \textit{co-hohh} are defined similarly. 

\begin{table}[h] %!tbp
 \centering
 {\small{
 \renewcommand{\arraystretch}{1.5}
 \begin{tabular}{p{1.5cm}|| p{5.1cm} ||  p{7.5cm} } 
   \hline
	& Program Clauses & Goals 
	\\ \hline \hline
	\emph{co-fohc} & $D \ ::= A^1 \mid G \supset D \mid  D\land D  \ \mid \forall Var \ D$ &  
	$G \ ::= \top \mid A^1 \mid G \land G \mid  G\lor G  \mid \exists Var \ G$ 
	\\ 
	\emph{co-hohc} & $D \ ::= A_r \mid G \supset D \mid  D\land D  \ \mid \forall Var \ D$ & $G \ ::= \top \mid A\hphantom{\scriptsize{1}} \mid G \land G \mid  G\lor G  \mid \exists Var \ G$ 
	\\ 
	\emph{co-fohh} & $D \ ::= A^1 \mid G \supset D \mid  D\land D  \ \mid \forall Var \ D$ & 
	$G \ ::= \top \mid A^1 \mid G \land G \mid  G\lor G  \mid \exists Var \ G \mid D \supset G\mid \forall Var\ G$ 
	\\ 
	\emph{co-hohh} & $D \ ::= A_r \mid G \supset D \mid  D\land D  \ \mid \forall Var \ D$ & $G \ ::= \top \mid A\hphantom{\scriptsize{1}} \mid G \land G \mid  G\lor G  \mid \exists Var \ G \mid D \supset G\mid \forall Var\ G$ 
	\\ \hline
\end{tabular}}
}
\caption{\textbf{D- and G-formulae}. $A$ and $A_r$ denote atoms and rigid atoms, respectively.  $A^1$ denote first-order atoms. In the setting of \emph{co-hohc}, $A$ and $A_r$ are from $\mathcal{U}^\Sigma_1$; in the setting of  \emph{co-hohh}, $A$ and $A_r$ are from $\mathcal{U}^\Sigma_2$. 
}
\label{tab:up}
\end{table}

\begin{figure}[h]
\centering
\small{
	$ \infer[\textsc{$\supset\!\! R$}]{\Sigma;P \longrightarrow D \supset G}{\Sigma;P, D \longrightarrow G} $ \hspace{1cm}
	$ \infer[\textsc{$\forall R$}]{\Sigma; P \longrightarrow \forall_{\!\tau} x\ G}{c:\tau,\Sigma; P \longrightarrow G\left[x:=c\right]}$ \hspace{1cm}
	$\infer[\textsc{$\exists R$}]{\Sigma;P \longrightarrow \exists_\tau x\ G}{\Sigma;P \longrightarrow G\left[x:=N\right] }$ 
		
	\vspace{7pt}	
	
	$ \infer[\textsc{$\lor R$}]{\Sigma;P \longrightarrow G_1 \lor G_2}{\Sigma;P \longrightarrow G_1 }\quad
	\infer[\textsc{$\lor R$}]{\Sigma;P \longrightarrow G_1 \lor G_2}{\Sigma;P \longrightarrow G_2 }$ \hspace{1cm} 
	$\infer[\textsc{$\land R$}]{\Sigma;P \longrightarrow G_1 \land G_2}{\Sigma;P \longrightarrow G_1 & \Sigma;P \longrightarrow G_2} $ 
	
	\vspace{5pt}
	 
	$\infer[\textsc{$\supset\! L$}]{\Sigma;P \stackrel{G \supset D}{\longrightarrow} A}{\Sigma;P \stackrel{D}{\longrightarrow} A & \Sigma;P \longrightarrow G }$ \hspace{1cm}
	$\infer[\textsc{$\land L$}]{\Sigma;P \stackrel{D_1 \land D_2}{\longrightarrow} A}{\Sigma;P \stackrel{D_1}{\longrightarrow} A}\quad
	\infer[\textsc{$\land L$}]{\Sigma;P \stackrel{D_1 \land D_2}{\longrightarrow} A}{\Sigma;P \stackrel{D_2}{\longrightarrow} A}$ \hspace{1cm}
	$\infer[\textsc{$\forall L$}]{\Sigma;P \stackrel{\forall_{\!\tau} x\ D}{\longrightarrow} A}{\Sigma;P \stackrel{D\left[x:=N\right]}{\longrightarrow} A }$ 
	
	\vspace{5pt}
	
		$ \infer[\textsc{decide}]{\Sigma; P \longrightarrow A}{\Sigma;P \stackrel{D}{\longrightarrow} A}$\hspace{1cm} 
		$ \infer[\textsc{initial}]{\Sigma; P \stackrel{A'}{\longrightarrow} A}{}$ \hspace{1cm}
		$\infer[\textsc{$\top R$}]{\Sigma;P \longrightarrow \top}{}\quad$ 
		}
	\caption[Uniform Proof]{\textbf{Uniform proof rules}.   
		\emph{Rule restrictions:} in $\exists R$ and $\forall L$, $N:\tau$ is a closed term on $\Sigma$. Moreover, 
		%\begin{itemize}
			%\item 
			if used in  \emph{co-fohc} or \emph{co-fohh}, then $N$ is first order;
			%\item 
			if used in  \emph{co-hohc}, then $N\in {\mathcal{U}_1^\Sigma}$; 
			%\item 
			if used in  \emph{co-hohh}, then  $N\in\mathcal{U}_2^\Sigma$.
		%\end{itemize}
		%\item 
		In $\forall R$,  $c:\tau\notin\Sigma$ ($c$ is also known as an \emph{eigenvariable}).
		%\item 
		In $\textsc{decide}$,  $D\in P$. 
		%\item 
		In the rule  \textsc{initial}, $A=_{\textit{fix}\beta} A'$.  
	%\end{itemize}
	}\label{fig: inductive uniform proof}
\end{figure}

\begin{figure}[h] 
\renewcommand{\arraystretch}{1.5}
\small{
\begin{center}
	$\infer[\textsc{co-fix}]{\Sigma;P \looparrowright M}{\Sigma;P,\langle M\rangle \longrightarrow \langle M\rangle}\quad$\begin{tabular}{p{1.5cm}  p{2.5cm}  p{1.5cm}  p{5cm} } 
	 \emph{co-fohc} & $M := A^1 \mid M\land M$ & \emph{co-fohh}  & $ M := A^1 \mid M \land M \mid M \supset M \mid \forall Var\ M$ \\
	%\hline
	\emph{co-hohc} & $M := A_r \mid M \land M$  & \emph{co-hohh} &  $M :=  A_r \mid M \land M \mid M \supset M \mid \forall Var\ M$
\end{tabular} 
\end{center}
}
	\caption[Co-fix rule]{\textbf{The coinductive fixed-point rule and syntax for core formulae.} \emph{Note:} In the upper sequent of \textsc{co-fix} rule, the left occurrence of $M$ is called a \emph{coinductive invariant}, and the right occurrence of $M$ is called a \emph{coinductive goal}.  The formula $M$ occurs on \emph{both} sides of the upper sequent in the \textsc{co-fix} rule, therefore $M$ must satisfy the syntax of both program clauses and goals. Formulae with such syntactic character as $M$ are called \emph{core formulae}~\cite{MN12}.
		   }\label{fig: general co-fix}
\end{figure}

\begin{figure}[h]
 	\small{
 	$\infer[\textsc{$\supset\!\! R$}\langle\rangle]{\Sigma;P \longrightarrow \langle M_1 \supset M_2\rangle}{\Sigma;P, \langle M_1 \rangle \longrightarrow \langle M_2\rangle} $
 	\ \ \
 	$\infer[\textsc{$\forall R$}\langle\rangle]{\Sigma; P \longrightarrow \langle \forall_{\!\tau} x\ M\rangle}{c:\tau,\Sigma; P \longrightarrow \langle M\left[x:=c\right]\rangle}$
   	 	\ \ \
		 	$\infer[\textsc{$\land R$}\langle\rangle]{\Sigma;P \longrightarrow \langle M_1 \land M_2\rangle}{\Sigma;P \longrightarrow \langle M_1\rangle & \Sigma;P \longrightarrow \langle M_2\rangle} $
 		
 		\vspace{5pt}
 			
 	 	$\infer[\textsc{$\supset\! L$}\langle\rangle]{\Sigma;P \stackrel{G \supset D}{\longrightarrow} \langle A\rangle }{\Sigma;P^* \stackrel{D}{\longrightarrow} A & \Sigma;P^* \longrightarrow  G }$
	\ \ \
 	 	$\infer[\textsc{$\land L$}\langle\rangle]{\Sigma;P \stackrel{D_1 \land D_2}{\longrightarrow} \langle A\rangle }{\Sigma;P \stackrel{D_1}{\longrightarrow} \langle A\rangle }$
 	%\hspace{1.5em}
 	 	\ \ \
	$\infer[\textsc{$\land L$}\langle\rangle]{\Sigma;P \stackrel{D_1 \land D_2}{\longrightarrow} \langle A\rangle }{\Sigma;P \stackrel{D_2}{\longrightarrow} \langle A\rangle }$
 	 	\ \ \
 	 	$\infer[\textsc{$\forall L$}\langle\rangle]{\Sigma;P \stackrel{\forall x\ D}{\longrightarrow} \langle A\rangle }{\Sigma;P \stackrel{D\left[x:= N\right]}{\longrightarrow} \langle A\rangle }$}
		 
		 \vspace{5pt}
		 	
		$\infer[\textsc{decide}\langle\rangle]{\Sigma; P \longrightarrow \langle A\rangle}{\Sigma;P \stackrel{D^*}{\longrightarrow} \langle A\rangle }$ 
		\ \ \
		$\infer[\textsc{initial}\langle\rangle]{\Sigma; P \stackrel{A'}{\longrightarrow} \langle A\rangle}{}$
 	\caption{\textbf{Rules for guarded coinductive goals.} \emph{Rule restrictions}: In $\textsc{decide}\langle\rangle$, $D^*$  must be a formula without  $\langle\rangle$ mark. In $\supset\! L \langle\rangle$,  $P^*$  results from erasing all $\langle\rangle$ marks in $P$. The restrictions for $\textsc{initial}\langle\rangle$, $\textsc{$\forall L$}\langle\rangle$ and $\textsc{$\forall R$}\langle\rangle$ are the same as for $\textsc{initial}$, $\forall L$ and $\forall R$ respectively. \emph{Note:}  Formulae added to the left-hand side by \textsc{co-fix} and $\supset\! R\langle\rangle$ are guarded, so that they are not selected by the $\textsc{decide}\langle\rangle$ rule for back-chaining with guarded atomic goals. The $\supset\! L\langle\rangle$ rule  frees all formulae from being guarded for each upper sequent, then rules in Figure~\ref{fig: inductive uniform proof} become applicable in further sequent reductions. }\label{fig: rules with angles}
 \end{figure}

\section{Discussion}
\label{sec: conl n dis}
% echo examples
Using coinductive uniform proofs, we can categorize infinite SLD-derivations, and we can uniformly and proof-theoretically formalize the coinductive reasoning performed by the two algorithms mentioned earlier. For instance,
% in the CoLP example, we need \textit{co-fohc} to express and prove the coinductive invariant {\footnotesize$(\textit{zeros}\ (\textit{fix}\ \lambda y\,.\,\textit{scons}\ 0 \ y))$}, and the root sequent for the uniform proof is $\Sigma_1;P_1\looparrowright (\textit{zeros}\ (\textit{fix}\ \lambda y\,.\,\textit{scons}\ 0 \ y))$; in the example for Fu et al's algorithm, we need \textit{co-fohh} to express and prove the coinductive invariant $\forall x\ (p\ x)$, and the root sequent is $\Sigma_2;P_2\looparrowright \forall x\ (p\ x)$;
to handle Example~\ref{ex:from}, we need \textit{co-hohh} extended with fixed point terms to express and prove the coinductive invariant $\forall x\ \textit{from}\ x\ (\textit{fr\_str}\ x)$, with the root sequent $\Sigma_5;\Gamma_5\looparrowright \forall x\ \textit{from}\ x\ (\textit{fr\_str}\ x)$.
%The full sequent proof for the last example is given in Appendix~\ref{sec: app}.  

We give the \emph{co-hohh} proof\footnote{We omit the subscript $5$ for $\Sigma,\Gamma$ in the proof.} for the sequent $\Sigma_5;\Gamma_5 \looparrowright \forall x (\mathit{from\ x\ (fr\_str\ x))}$. Note that $fr\_str$ is defined in Section \ref{sec: languages and rules}, $CH$ abbreviates the coinductive hypothesis $\mathit{\forall x (from\ x\ (fr\_str\ x))}$,  $Z$ is an arbitrary eigenvariable,   and the step marked by $\checkmark$ indicates involvement of the relation \[\mathit{from}\ Z\ (\textit{scons}(Z, (fr\_str\ (s \ Z)))) 	=_{\textit{fix}\beta} \mathit{from\ Z\ (fr\_str\ Z)}\] 
The two $\forall L\langle \rangle$ steps involve the substitutions $x:= Z, y:=(fr\_str\ (s \ Z))$. The $\forall L$ step involves the substitution $x:= s\ Z$.
\begin{displaymath}
	\resizebox{\textwidth}{!}{
	\infer[\textsc{co-fix}]		
	{
		\Sigma;\Gamma \looparrowright \mathit{\forall x (from\ x\ (fr\_str\ x))}
	}
	{	
		\infer[\forall R \langle \rangle]
		{\Sigma;\Gamma,\langle CH\rangle\longrightarrow\mathit{\langle \forall x (from\ x\ (fr\_str\ x)) \rangle }}
		{
			\infer[\textsc{decide}\langle \rangle]	
			{
				Z,\Sigma;\Gamma,\langle CH\rangle\longrightarrow\mathit{\langle  from\ Z\ (fr\_str\ Z) \rangle}
			}
			{	
				\infer[\forall L \langle \rangle\left(\text{2 times}\right)]
				{		
					Z,\Sigma;\Gamma,\langle CH\rangle\stackrel{\forall xy\ \textit{from}\ (s\ x)\ y \supset  \textit{from}\ x\ (\textit{scons}\ x \ y)}{\longrightarrow}\mathit{\langle from\ Z\ (fr\_str\ Z) \rangle}  					
				}				
				{	
					\infer[\supset L \langle \rangle]	
					{
						Z,\Sigma;\Gamma,\langle CH\rangle\stackrel{\mathit{from\ (s\ Z) (fr\_str\ (s\ Z))\  \;\supset\; from\ Z\ (\textit{scons}\ Z\ (fr\_str\ (s \ Z))) }}{\longrightarrow}\mathit{\langle from\ Z\ (fr\_str\ Z) \rangle}	
					}
					{	
						\infer[\textsc{initial}\checkmark]	
						{
							Z,\Sigma;\Gamma,CH\stackrel{\mathit{from}\ Z\ (\textit{scons}\ Z\ (fr\_str\ (s \ Z))) }{\longrightarrow}\mathit{from\ Z\ (fr\_str\ Z)}
						}
						{}                 
						&
						\infer[\textsc{decide}]
						{
							Z,\Sigma;\Gamma,CH\longrightarrow \mathit{from\  (s\ Z)\  (fr\_str\ (s \ Z))}	}
						{ 
							\infer[\forall L]
							{
							Z,\Sigma;\Gamma,CH \stackrel{CH}{\longrightarrow} \mathit{from\  (s\ Z)\  (fr\_str\ (s \ Z))}
							}
							{
								\infer[\textsc{initial}]
								{Z,\Sigma;\Gamma,CH \stackrel{\mathit{from\ (s \ Z)\ (fr\_str\ (s \ Z))}}{\longrightarrow} \mathit{from\  (s\ Z)\  (fr\_str\ (s \ Z))}}
								{}
							}	
						}
					}
				}
			}
		}			
	}
} % resizebox
	\end{displaymath}

        Given this proof, we can obtain the proof for $\mathit{from\ 0\ (fr\_str\ 0)}$ as a corollary. This is exactly the goal we were not able to acheive in Example~\ref{ex:from} by loop detection.

        The fact that the \textsc{co-fix} rule can only be applied once and as the first step in a proof,  is a simplification that helps to highlight the basic coinductive argument performed by the coinductive uniform proofs. The absence of nested coinduction in the meta-theory can be mitigated by allowing using the already proven coinductive invariants as lemmas to prove further coinductive conclusions.  

\section{Future Work}

We omit technical details of the proof of soundness of coinductive uniform proofs w.r.t greatest complete Herbrand models. Intuitively, the  proof proceeds by defining a scheme by which we can reconstruct a corresponding non-terminating derivation, 
and then showing that the proofs are sound w.r.t greatest complete Herbrand models. However, in contrast with CoLP, % rather than simple ``copy-and-paste'' of the template as 
the reconstruction is generally more complicated and involves 
\begin{itemize}
	\item 
          %\emph{(i)}
          a construction of a function that generates countably many different substitution instances for the derivation scheme, and 
	\item 
          %\emph{(ii)}
          showing that these instances can be composed in a certain way 
	%with necessary  ``copy-and-paste'' 
	in order to restore the full infinite derivation.
\end{itemize}
The proof is constructive, and in addition uses a coinductive proof principle when showing correspondence of the derivation schemes to greatest complete Herbrand model construction.

Now that we have a sound framework for automated coinductive proof construction, the practical problem is to formulate heuristics  that
can find suitable coinductive invariants to prove. It can be shown that CoLP method in fact finds coinductive invariants
expressed and proven in \emph{co-fohc} (with and without fixed point terms). The method presented in~\cite{FKS15} formulates coinductive invariants
in \emph{co-fohh} (without fixed point terms). The current work is on the way to generalise these methods to other logics.

% ~\cite{Dowek2011PCF}
% A similar approach to guarding coinductive proofs has recently been suggested independently (and in a different calculus) by Basold~\cite{B18}.

\bibliographystyle{eptcs}
\bibliography{katya2}

%\appendix
%\input{syntax}
%\input{fixp}
%\input{fp}
%\input{trees}
%\input{lattice}
%\input{sproof}
%\input{Coq}
%\input{proof-reg}
%\input{proof-irreg} 
\end{document}